%% file: main.tex
\documentclass{ifacconf}

\usepackage{graphicx,mathtools,amsfonts}      
\usepackage{natbib}        

\usepackage{pgfplots}
\usepgfplotslibrary{fillbetween}
    \pgfplotsset{compat=newest}
     \usetikzlibrary{external}
    \usepgfplotslibrary{statistics}

\newtheorem{theorem}{Theorem}
\newtheorem{proposition}{Proposition}
\DeclareMathOperator{\vspan}{span}
\DeclareMathOperator{\diag}{diag}
\newcommand{\ifacqed}{\hfill\vphantom{l}\qed}
\begin{document}
\begin{frontmatter}

\title{On Convergent Dynamic Mode Decomposition and its Equivalence with Occupation Kernel Regression\thanksref{footnoteinfo}} 

\thanks[footnoteinfo]{This work was supported in part by the Air Force Office of Scientific Research (AFOSR) under Contracts FA9550-20-1-0127 and FA9550-21-1-0134, and in part by the National Science Foundation (NSF) under Awards 2027999 and 2027976. Any opinions, findings, or conclusions in this paper are those of the author(s) and do not necessarily reflect the views of the sponsoring agencies.}

\author[First]{Moad Abudia} 
\author[Second]{Joel Rosenfeld} 
\author[First]{Rushikesh Kamalapurkar}

\address[First]{Oklahoma State University, 
   Stillwater, OK, USA (e-mail: \{abudia,rushikesh.kamalapurkar\}@okstate.edu).}
\address[Second]{University of South Florida, 
   Tampa, FL, USA (e-mail: rosenfeldj@usf.edu)}

\begin{abstract}
This paper presents a new technique for norm-convergent dynamic mode decomposition of deterministic systems. The developed method utilizes recent results on singular dynamic mode decomposition where it is shown that by appropriate selection of domain and range Hilbert spaces, the Liouville operator (also known as the Koopman generator) can be made to be compact. In this paper, it is shown that by selecting appropriate collections of finite basis functions in the domain and the range, a novel finite-rank representation of the Liouville operator may be obtained. It is also shown that the model resulting from dynamic mode decomposition of the finite-rank representation is closely related to regularized regression using the so-called occupation kernels as basis functions.
\end{abstract}


\end{frontmatter}

\section{Introduction}
Dynamic mode decomposition (DMD) methods are data analysis methods that aim to generate a finite-rank representation of a transfer operator corresponding to a nonlinear dynamical system using time series measurements \citep{SCC.Kutz.Brunton.ea2016,SCC.Budisic.Mohr.ea2012,SCC.Mezic2005,SCC.Korda.Mezic2018}. 
The convergence of the finite-rank representations to the true transfer operator (the Koopman operator) has been established in results such as \cite{SCC.Korda.Mezic2018}, but only with respect to the strong operator topology (SOT). Convergence in SOT does not guarantee the convergence of the spectrum \citep{SCC.Pedersen2012}, and therefore the corresponding dynamic mode decomposition (DMD) algorithms, which rely on spectrum of the operator, are not guaranteed to converge.

In this paper, the above limitations are addressed by removing Koopman operators from the analysis in favor of Liouville operators (known as Koopman generators in special cases). These operators are shown to be compact provided their domains and ranges are selected appropriately. The result is a norm convergent finite-rank representation which significantly improves upon the aforementioned SOT convergent results.

There have been several attempts to provide compact operators for DMD. The approaches in \cite{SCC.Das.Giannakis.ea2021} and \cite{SCC.Rosenfeld.Kamalapurkar.ea2022} find compact operators through the multiplication of an auxiliary operator against Koopman and Liouville operators, respectively. However, the resultant operators only approximately correspond to the dynamics in question, and as such, the resulting DMD methods, while useful, are heuristic. 

The approach in this paper generates compact Liouville operators that truly correspond to the given continuous-time dynamics. Such operators were shown to exist for a large class of nonlinear systems in \cite{SCC.Rosenfeld.Kamalapurkar2023a}, where norm-convergent finite-rank representations were also derived via the adjoint of the operators. In this paper, we develop a finite-rank representation that approximates the Liouville operator directly, rather than through its adjoint. The direct approximation makes the resulting DMD algorithm numerically efficient and conceptually easier. Interestingly, the resulting model is equivalent to the occupation kernel regression (OKR) model developed in \cite{SCC.Li.Rosenfeld2020}.

OKR is a generalization of kernel ridge regression \cite{SCC.Zhdanov.Kalnishkan2013} where the loss function is defined in terms of inner products of components of the vector field that models the dynamics of the system and trajectory-dependent functions in a reproducing kernel Hilbert space (RKHS) called occupation kernels that represent integration along the trajectory. A Representer theorem is used to construct an approximation of the vector field as a linear combination of occupation kernels. Since the DMD model developed in this paper is seen to be identical to the OKR model without regularization, convergence results derived in this paper are also applicable to OKR, provided the regularization parameter is set to zero. To facilitate the discussion, the following section recalls a few important characteristics of DMD methods and RKHSs.

\section{Reproducing Kernel Hilbert Spaces and Dynamic Mode Decomposition} \label{sec:rkhs}
An RKHS, $H$, over a compact set $X$, is a space of functions from $X$ to $\mathbb{R}$ such that the evaluation functional $E_x:H\to\mathbb{R}$, defined as $E_x g := g(x)$, is bounded for every $x \in X$. By the Riesz representation theorem \cite[Theorem 13.32]{SCC.Roman2008}, for each $x \in X$ there exists a function $K_x \in H$ such that $\langle f, K_x \rangle_H = f(x)$ for all $f \in H$. The function $K_x$ is called the kernel function centered at $x$, and the function $K(x,y) := \langle K_y, K_x \rangle_H$ is called the reproducing kernel of $H$ \cite[Chapter 4]{SCC.Steinwart.Christmann2008}.

A symmetric function $K:X\times X\to\mathbb{R}$ is called a positive semidefinite kernel if for every integer $M>0$ and every finite collection of points $\{ x_1, \ldots, x_M \} \subset X$, the Gram matrix $( K(x_i,x_j) )_{i,j=1}^M$ is positive semidefinite. By the Aronszajn-Moore theorem \citep{SCC.Aronszajn1950}, given any positive semidefinite kernel $K$, there exists a unique RKHS $H$ such that $K$ is the reproducing kernel of $H$.

The motivation in DMD is to compute an invariant subspace of a transfer operator $A_f$ that models the evolution of test functions along the trajectories of a dynamical system $\dot{x} = f(x)$. The transfer operator maps a test function $g$ to its time derivative $ \nabla g\cdot f$. The subspace is typically constructed as the span of eigenfunctions of the operator. While such transfer operators over RKHSs may not admit point spectra, DMD methods aim to construct a finite-rank representation of the transfer operator and to leverage the spectrum of the approximating operator for modeling. 

The objective is to find functions for which
\begin{equation}\label{eq:eig_goal} |A_f \phi(x) - \lambda \phi(x)| < \epsilon \end{equation} for some $\lambda$ and some small positive $\epsilon$ and all $x$ within a domain of interest. Since norm convergence in a RKHS of continuous functions yields uniform convergence over compact sets \citep{SCC.Steinwart.Christmann2008,SCC.Wendland2004}, it is sufficient to satisfy $\|A_f \phi - \lambda \phi\|_H < \epsilon$. In turn, if a finite-rank approximation of $A_f$, call it $\tilde A_f$, is close enough, it is sufficient to satisfy $\|A_f - \tilde A_f\| < \epsilon$, and the rest follows as
\begin{gather*}
    | A_f \phi(x) - \lambda \phi(x) | < C \| A_f \phi - \lambda \phi \|_H\\
    C \| A_f \phi - \tilde A_f \phi \|_H < C \| A_f - \tilde A_f \|_H < C \epsilon,
\end{gather*}
where $C$ is a positive constant that depends on the domain of interest and the kernel function, and the function $\phi$ is assumed to be normalized.
        
A convergent approximation of the spectrum of the transfer operator using the spectrum of finite-rank operators requires compactness and convergence in norm, which motivates the following section.

\section{Compact Liouville Operators and Occupation Kernels} \label{sec:compact_liouville}
This section, included here for completeness, recalls the results from \cite{SCC.Rosenfeld.Kamalapurkar2023a} pertaining to the existence of compact Liouville operators, given formally as $A_f g(x) = \nabla g(x) f(x)$. Compactness is achieved through the consideration of differing spaces for the domain and the range of the operator. We emphasize that compact Liouville operators are not restricted to these particular pairs of functions spaces. This section only provides examples demonstrating the existence of such operators, thereby validating the approach in the sequel.

\subsection{Liouville Operators on Real Bergmann-Fock Spaces}\label{sec:severalvariables}
The exponential dot product kernel, with parameter $\mu > 0$, is given as $K(x,y) = \exp\left(\frac{x^\top y}{\mu}\right)$. In the single variable case, the native space for this kernel is the restriction of the Bergmann-Fock space to real numbers, denoted by $F^2_{\mu}\left(\mathbb{R}\right)$. This space consists of the set of polynomials of the form $h(x) = \sum_{k=0}^\infty a_k x^k$, where the coefficients satisfy $\sum_{k=0}^\infty \left\vert a_k\right\vert^2 \mu^k k! < \infty$, and the norm is given by $ \left\Vert h \right\Vert_{\mu}^2 = \sum_{k=0}^\infty \left\vert a_k\right\vert^2 \mu^k k! $. Extension of this definition to the multivariable case yields the space $F^2_{\mu}\left(\mathbb{R}^n\right)$ where the collection of monomials, $x^{\alpha} \frac{\mu^{|\alpha|}}{\sqrt{\alpha!}}$, with multi-indices $\alpha \in \mathbb{N}^n$ forms an orthonormal basis, where for $\alpha \in \mathbb{N}^n$, $\alpha! = \prod_{i=1}^n \alpha_i !$, $|\alpha| = \sum_{i=1}^n \alpha_i$, and $x^\alpha = \prod_{i=1}^n x_i^{\alpha_i}$. In this setting, differential operators from $F^2_{\mu_1}(\mathbb{R}^n)$ to $F^2_{\mu_2}(\mathbb{R}^n)$ can be shown to be compact provided  $\mu_2 < \mu_1$.
\begin{proposition}[\cite{SCC.Rosenfeld.Kamalapurkar2023a}]\label{prop:differential_compact}
    If $\mu_2 < \mu_1$, then the differential operators $\frac{\partial}{\partial x_i}  : F^2_{\mu_1}(\mathbb{R}^n) \to F^2_{\mu_2}(\mathbb{R}^n)$, are compact for $i=1,\ldots,n$. 
\end{proposition}
Multiplication operators can be shown to be bounded provided that their symbols are polynomial.
\begin{proposition}[\cite{SCC.Rosenfeld.Kamalapurkar2023a}]\label{prop:multiplication_bounded}
    If $\mu_2 < \mu_1$, then for any polynomial function $p:\mathbb{R}^n \to \mathbb{R} $, the multiplication operator $M_{p} : F^2_{\mu_1}(\mathbb{R}^n) \to F^2_{\mu_2}(\mathbb{R}^n)$, defined as $\left[M_p h\right](x) = p(x)h(x)$, is bounded.
\end{proposition}
Boundedness of Liouville operators with polynomial symbols then follows trivially from the above two results.
\begin{theorem}[\cite{SCC.Rosenfeld.Kamalapurkar2023a}]\label{thm:compactness}
    If $\mu_2 < \mu_1$ and if $f:\mathbb{R}^n \to \mathbb{R}^n$ is a component-wise polynomial function, then the Liouville operator $A_f : F^2_{\mu_1}(\mathbb{R}^n) \to F^2_{\mu_2}(\mathbb{R}^n)$ defined as $A_f g = \nabla g \cdot f$ is a compact operator. 
\end{theorem}
\subsection{Occupation Kernels and Liouville Operators}
Let $H$ be an RKHS over a compact set $X \in \mathbb{R}^n$ consisting of continuous functions and let $K$ be the reproducing kernel of $H$. Given a continuous signal $\gamma:[0,T] \to X$, the linear functional $g \mapsto \int_0^{T} g(\gamma(t)) \mathrm{d}t$ is bounded. Hence, by the Riesz representation theorem \cite[Theorem 13.32]{SCC.Roman2008}, there exists a function $\Gamma_{\gamma} \in H$ such that $\langle g, \Gamma_{\gamma} \rangle_{H} = \int_0^T g(\gamma(t)) \mathrm{d}t$ for all $g \in H$. The function $\Gamma_{\gamma}$ is called the occupation kernel corresponding to $\gamma$ in $H$. These occupation kernels were first introduced in \cite{SCC.Rosenfeld.Kamalapurkar.ea2019a,SCC.Rosenfeld.Russo.ea2024}. 

Occupation kernels corresponding to trajectories of a dynamical system have a useful relationship with adjoints of Liouville operators and the reproducing kernels of the underlying RKHSs.
\begin{proposition}[\cite{SCC.Rosenfeld.Kamalapurkar.ea2022}]\label{prop:AdjointOCCRelation}
    If $\gamma:[0,T]\to X$ is a solution of $\dot{x} = f(x)$ for some locally Lipschitz continuous function $f:X\to\mathbb{R}^n$, then $A_f^* \Gamma _{\gamma } = K(\cdot , \gamma (T)) - K(\cdot ,\gamma (0))$.
\end{proposition}
\begin{proposition}[\cite{SCC.Rosenfeld.Kamalapurkar.ea2019}]
    Given any continuous function $\gamma:[0,T]\to X$, the occupation kernel corresponding to $\gamma$ in $H$ may be expressed as ${\Gamma _\gamma }(x) = \int_0^T K (x,\gamma (t))\mathrm{d}t$. As a consequence, given two functions $\gamma_i:[0,T_i]\to X$ and $\gamma_j:[0,T_j]\to X$  \begin{equation}
        \left\langle\Gamma_{\gamma_{j}}, \Gamma_{\gamma_{i}}\right\rangle_{H}=\int_{0}^{T_{i}} \int_{0}^{T_{j}} K\left(\gamma_{i}(\tau), \gamma_{j}(t)\right) \mathrm{d} t \mathrm{d} \tau.
    \end{equation}
\end{proposition}
\section{Finite-rank Representation of the Liouville Operator}\label{sec:finite_rank}
This section provides a novel finite-rank representation of the Liouville operator $A_{f}$ and subsequently, a novel approach to obtain the dynamic modes of the underlying dynamical system. Let $H_d$ and $H_r$ be RKHSs with reproducing kernels $K_d$ and $K_r$, respectively. The RKHS $H_d$ is used as the domain of the Liouville operator and the RKHS $H_r$ is used as the range. In what follows, finite collections of vectors $d^M \subset H_d$ and $r^M \subset H_r$ are selected to establish the needed finite-rank representation of the Liouville operator. 

Since the adjoint of $ A_{f} $ maps occupation kernels to kernel differences (Proposition \ref{prop:AdjointOCCRelation}), the span of the collection of kernel differences
\begin{equation}
    d^M = \left\{K_d(\cdot,\gamma_{i}(T_i)) - K_d(\cdot,\gamma_{i}(0))\right\}_{i=1}^M\subset H_{d}
\end{equation}
are selected to be the domain of  $ A_{f} $.  The corresponding Gram matrix is denoted by $G_{d^M} = \left(\left\langle d_i,d_j\right\rangle_{H_d}\right)_{i,j=1}^M$. The result of $A_f$ operating on functions in $\vspan{d^M}$ is projected onto the span of the collection of occupation kernels
\begin{equation}
   \vspan{r^M}  =  \vspan{\left\{\Gamma_{\gamma_{i}}\right\}_{i=1}^M}\subset H_{r}.
\end{equation}
The corresponding Gram matrix is denoted by $G_{r^M} = \left(\left\langle r_i,r_j\right\rangle_{H_r}\right)_{i,j=1}^M$. A rank-$M$ representation of the operator $A_{f}$ is then given by  $P_{r^M} A_{f} P_{d^M}:H_d\to\vspan r^M$, where $P_{r^M}$ and $P_{d^M}$ denote projection operators onto $\vspan{r^M}$ and $\vspan{d^M}$, respectively. Under the compactness assumptions and given rich enough data so that the spans of $ \{d_i\}_{i=1}^\infty $ and $\{r_i\}_{i=1}^\infty$ are dense in $H_{d}$ and $H_{r}$, respectively, the sequence of finite-rank operators $\{P_{r^M} A_{f} P_{d^M}\}_{M=1}^{\infty}$ can be shown to converge, in the norm topology, to $A_{f}$. 
To facilitate the proof of convergence, we recall the following result from \cite{SCC.Rosenfeld.Kamalapurkar.ea2022}.
\begin{proposition}[\cite{SCC.Rosenfeld.Kamalapurkar.ea2022}]\label{lem:technical-lemma}
    Let $H_d$ and $H_r$ be RKHSs defined on $X\subset \mathbb{R}^n$ and let $A_N:H_d\to H_r$ be a finite-rank operator with rank $N$. If $\vspan\{d_i\}_{i=1}^{\infty} \subset H_d$ is dense in $H_d$ and $\vspan\{r_i\}_{i=1}^{\infty} \subset H_r$ is dense in $H_r$, then for all $\epsilon > 0 $, there exists $M(N)\in\mathbb{N}$ such that for all $i\geq M(N)$ and $h\in H_d$, $\left\Vert A_N h - A_N P_{d^i} h\right\Vert_{H_r} \leq \epsilon \left\Vert h \right\Vert_{H_d}$ and $\left\Vert A_N h - P_{r^i} A_N h\right\Vert_{H_r} \leq \epsilon \left\Vert h \right\Vert_{H_d}$. 
\end{proposition}
The convergence result for Liouville operators on Bergmann-Fock spaces restricted to the set of real numbers follows from the following more general result.
\begin{proposition}\label{prop:convergence}
    If $A:H_d\to H_r$ is a compact operator, and spans of the collections $ \{d_i\}_{i=1}^\infty $ and $\{r_i\}_{i=1}^\infty$ are dense in $H_{d}$ and $H_{r}$, respectively, then $\lim_{M\to\infty}\left\Vert A - P_{r^M} A P_{d^M} \right\Vert_{H_d}^{H_r} = 0$, where $\left\Vert\cdot\right\Vert_{H_d}^{H_r}$ denotes the operator norm of operators from $H_d$ to $H_r$.
\end{proposition}
\begin{pf}
    Let $\{A_N\}_{N=1}^\infty$ be a sequence of rank-$N$ operators converging, in norm, to $A$. For an arbitrary $h\in H_d$, 
    \begin{multline*}
        \left\Vert Ah - P_{r^M} A P_{d^M} h\right\Vert_{H_r}
        \leq \left\Vert Ah - A_{N}h  \right\Vert_{H_r} \\
        + \left\Vert A_{N}h - A_{N}P_{d^M}h \right\Vert_{H_r} 
        +\left\Vert A_{N}P_{d^M}h - P_{r^M}A_{N}P_{d^M}h \right\Vert_{H_r}\\ + \left\Vert P_{r^M}A_{N}P_{d^M}h - P_{r^M} A P_{d^M}h \right\Vert_{H_r}.
    \end{multline*}
    Using the fact that $A_N$, $P_{r^M}A_{N}$, and $A_{N}P_{d^M}$ are all finite-rank operators and the fact that the projection operator is bounded with norm bound 1, Proposition \ref{lem:technical-lemma}, can be used to conclude that for all $ \epsilon > 0 $, there exists $M(N)\in\mathbb{N}$ such that for all $i\geq M(N)$
    \begin{multline*}
        \left\Vert Ah - P_{r^i} A P_{d^i} h\right\Vert_{H_r}
        \leq \left\Vert (A - A_{N})h  \right\Vert_{H_r} 
        + 2\epsilon\left\Vert h \right\Vert_{H_d} \\
         + \left\Vert (A_{N} - A)P_{d^i}h \right\Vert_{H_r}.
    \end{multline*}
    Since $A_{N}$ converges to $A$ in norm, given $\epsilon > 0$, there exists $N\in \mathbb{N}$ such that for all $j\geq N$, and  $g\in H_d$ $\left\Vert Ag - A_{j}g  \right\Vert_{H} \leq \epsilon \left\Vert g  \right\Vert_{H_d} $. Thus, for all  $j\geq N$ and  $i\geq M(j)$, $ \left\Vert Ah - P_{r^i} A P_{d^i} h\right\Vert_{H_r} \leq 4 \epsilon \left\Vert h \right\Vert_{H_d}$.\ifacqed
\end{pf}

\subsection{Matrix Representation of the Finite-rank Operator}
 To formulate a matrix representation of the finite-rank operator $P_{r^M} A_{f} P_{d^M} $, the operator is restricted to $\vspan d^M$ to yield the finite-rank operator $P_{r^M} A_{f}|_{d^M}:\vspan{d^M}\to\vspan{r^M}$. The resulting matrix representation is denoted by $[A_{f}]_d^r$. For brevity of exposition, the superscript $M$ is suppressed hereafter and $d$ and $r$ are interpreted as $M-$dimensional vectors.
\begin{theorem}
    Let $g = a^\top r\in \vspan{r}$ and $ h = \delta^\top d\in \vspan{d} $ be functions with coefficients $ a\in\mathbb{R}^M$ and $\delta\in\mathbb{R}^M $, respectively. If $g = P_r A_{f}|_d h$, then $ a = G_r^{+} G_d \delta $. That is, $[A_{f}]_d^r \coloneqq G_r^{+} G_d$, where $G_r^{+}$ denotes the Moore-Penrose pseudoinverse of $G_r$, is a matrix representation of $P_r A_{f}|_d$.
\end{theorem}
\begin{pf}
    Since $g = P_r A_{f}|_d h = a^\top r$, the coefficients $a$ solve
    \begin{equation*}
        G_r a  
        =\begin{bmatrix}
            \left\langle A_f\delta^\top d,r_1 \right\rangle_{H_r}\\\vdots\\\left\langle A_f\delta^\top d,r_M \right\rangle_{H_r}
        \end{bmatrix}=\begin{bmatrix}
            \left\langle \delta^\top d,A_f^*r_1 \right\rangle_{H_d}\\\vdots\\\left\langle \delta^\top d,A_f^*r_M \right\rangle_{H_d}
        \end{bmatrix}
    \end{equation*}
    Proposition \ref{prop:AdjointOCCRelation} implies that $A_f^*r_i = d_i$ for all $i=1,\cdots,M$, and as such, $G_r a = G_d \delta$.
    Since $ a = G_r^{+} G_d \delta $ is a solution of $G_r a = G_d \delta$, we have, $a = G_r^{+} G_d \delta$, and as a result, $G_r^{+} G_d$ is a matrix representation of $P_r A_{f}|_d$. \ifacqed
\end{pf}
In the following section, the matrix representation $[A_{f}]_d^r$ is used to construct a data-driven representation of the singular values and the left and right singular functions of the operator $P_r A_{f}\vert_d$. 

\subsection{Singular Functions of the Finite-rank Operator}
The tuples $\{(\sigma_i,\phi_i,\psi_i)\}_{i=1}^M$, with $\sigma_i\in \mathbb{R}$, $\phi_i\in H_{d}$, and $\psi_i\in H_{r}$, are singular values, left singular vectors, and right singular vectors of $P_r A_{f}|_d$, respectively, if $\forall h\in\vspan{d}$, $ P_r A_{f} h = \sum_{i=1}^M \sigma_i \psi_i \left\langle h,\phi_i\right\rangle_{H_{d}}$.
The following proposition states that the SVD of $P_r A_{f}|_d$ can be computed using matrices in the matrix representation $[A_{f}]_d^r$ developed in the previous section.
\begin{theorem}\label{thm:svd}
    If $ (W,\Sigma,V) $ is the SVD of $G_r^{+}$ with $W = \begin{bmatrix} w_1,&\ldots,&w_M \end{bmatrix}$, $V = \begin{bmatrix} v_1,&\ldots,&v_M \end{bmatrix}$, and $\Sigma = \diag\left(\begin{bmatrix} \sigma_1,&\ldots,&\sigma_M \end{bmatrix}\right)$, then for all $i=1,\ldots,M$, $\sigma_i$ are singular values of of $P_r A_{f}|_d$ with left singular functions $\phi_i := v_i^\top d$ and right singular functions $\psi_i := w_i^\top r$, respectively.
\end{theorem}
\begin{pf}
    Let $\phi_i = v_i^\top d$ and $\psi_i = w_i^\top r$ and $h = \delta^\top d$. Then, 
    \begin{multline*}
        P_r A_{f} h = \sum_{i=1}^M \sigma_i \psi_i \left\langle h,\phi_i\right\rangle_{H_{d}}
        \iff \\
        P_r A_{f} \delta^\top d = \sum_{i=1}^M \sigma_i w_i^\top r \left\langle \delta^\top d, v_i^\top d\right\rangle_{H_{d}}
    \end{multline*}
    Using the finite-rank representation, the collection $\{(\sigma_i,\phi_i,\psi_i)\}_{i=1}^M$, is a SVD of $P_r A_{f}|_d$, if for all $\delta\in\mathbb{R}^M$,
    \begin{equation}\label{eq:suff_cond_SVD}
        \left(G_r^{+} G_d \delta\right)^\top r = \left(\sum_{i=1}^M \sigma_i \left\langle \delta^\top d, v_i^\top d\right\rangle_{H_{d}} w_i^\top \right)r.
    \end{equation}
    Simple matrix manipulations yield the chain of implications
    \begin{multline*}
       \eqref{eq:suff_cond_SVD}\impliedby \forall \delta\in\mathbb{R}^M,G_r^{+} G_d \delta = \sum_{i=1}^M \sigma_i \left\langle \delta^\top d, v_i^\top d\right\rangle_{H_{d}} w_i\\
        \iff \forall \delta\in\mathbb{R}^M,G_r^{+} G_d \delta 
        = \sum_{i=1}^M \sigma_i  \left(w_i v_i^\top G_d\right) \delta\\
        \impliedby G_r^{+} G_d = \sum_{i=1}^M \sigma_i  \left(w_i v_i^\top \right)G_d\\
        \impliedby G_r^{+} = \sum_{i=1}^M \sigma_i w_i v_i^\top = W\Sigma V^\top,
    \end{multline*}
    which proves the proposition.\ifacqed
\end{pf}
In the following section, the singular values and the left and right singular vectors are used to generate a data-driven model via a method termed singular Liouville dynamic mode decomposition (SLDMD).

\section{The SLDMD Algorithm}\label{sec:SLDMD}
Let the identity observable $(h_{\mathrm{id}})_j$ be defined as $(h_{\mathrm{id}})_j(x) = x_j$, where $x_j$ is the $j-$th component of $x$.

\begin{theorem}
    If $(h_{\mathrm{id}})_j \in H_{d}$ for $j=1,\cdots,n$, then
    \begin{equation}\label{eq:convergent_model}
        \hat{f}_{M} (x) := \begin{bmatrix}
            [P_{r} A_{f} P_{d} (h_{\mathrm{id}})_1] (x)\\
            \vdots\\
            [P_{r} A_{f} P_{d} (h_{\mathrm{id}})_n] (x)
        \end{bmatrix} = DG_r^{+} r(x)
    \end{equation}
    where $D \coloneqq \left(\left(\gamma_{j}(T_j)\right)_i - \left(\gamma_{j}(0)\right)_i\right)_{i,j=1}^{n,M} $ and $(\gamma_j(\cdot))_i$ denotes the $i-$th component of $\gamma_j(\cdot)$.
\end{theorem}
\begin{pf} Using Theorem \ref{thm:svd} and the definition of singular values and singular functions of $P_{r} A_{f}|_d$,
\begin{equation}
    \hat{f}_{M}(x) = 
    \sum_{i=1}^M \sigma_i \xi_i w_i^\top r (x) = \xi\Sigma W^\top r(x),
\end{equation}
where $\xi_i \coloneqq \begin{bmatrix}
    \left\langle P_d (h_{\mathrm{id}})_1,\phi_i\right\rangle_{H_{d}}&\ldots,\left\langle P_d (h_{\mathrm{id}})_n,\phi_i\right\rangle_{H_{d}}
\end{bmatrix}^\top$ and $\xi := \begin{bmatrix}
    \xi_1&\ldots&\xi_M
\end{bmatrix}$.

The modes $\xi$ can be computed using $\phi_i = v_i^\top d$ as
\begin{multline*}
    \xi = \begin{bmatrix}
    \left\langle P_d (h_{\mathrm{id}})_1,v_1^\top d\right\rangle_{H_{d}},&\ldots,&\left\langle P_d (h_{\mathrm{id}})_1,v_M^\top d\right\rangle_{H_{d}}\\
    \vdots & \ddots & \vdots\\
    \left\langle P_d (h_{\mathrm{id}})_n,v_1^\top d\right\rangle_{H_{d}},&\ldots,&\left\langle P_d (h_{\mathrm{id}})_n,v_M^\top d\right\rangle_{H_{d}}
    \end{bmatrix}
    \\= \begin{bmatrix}
        \left\langle \delta_1^\top d,d_1\right\rangle_{H_{d}},&\ldots,&\left\langle \delta_1^\top d,d_M\right\rangle_{H_{d}}\\
        \vdots & \ddots & \vdots\\
    \left\langle \delta_n^\top d,d_1\right\rangle_{H_{d}},&\ldots,&\left\langle \delta_n^\top d,d_M\right\rangle_{H_{d}}
    \end{bmatrix} V
    = \delta^\top G_d V,
\end{multline*}
where $\delta \coloneqq  \begin{bmatrix} \delta_1, &\ldots, &\delta_n \end{bmatrix}$. Using the reproducing property of the reproducing kernel of $H_d$, the coefficients $\delta_i$ in the projection of $(h_{\mathrm{id}})_i$ onto $d$ solve the system of linear equations
\begin{equation*}
    G_d\delta_i = \begin{bmatrix}
        \left\langle\left(h_{\mathrm{id}}\right)_i,d_1\right\rangle_{H_{d}}\\ \vdots \\ \left\langle\left(h_{\mathrm{id}}\right)_i,d_M\right\rangle_{H_{d}}
    \end{bmatrix} = \begin{bmatrix}
        \left(\gamma_{1}(T_1)\right)_i - \left(\gamma_{1}(0)\right)_i\\ \vdots \\ \left(\gamma_{M}(T_M)\right)_i - \left(\gamma_{M}(0)\right)_i
    \end{bmatrix}.
\end{equation*}
Letting $D \coloneqq \left(\left(\gamma_{j}(T_j)\right)_i - \left(\gamma_{j}(0)\right)_i\right)_{i,j=1}^{n,M} $ it can be concluded that $ \delta^\top G_d = D $. Finally, the modes $\xi$ are given by $\xi = D V$ and 
\begin{equation*}
    \hat{f}_{M}(x) = D V\Sigma W^\top r(x) = DG_r^{+} r(x).\ifacqed
\end{equation*}
\end{pf}
Using $\hat{f}_{M}$ as a rank$-M$ estimate of $f$, the estimated system model is of the form $\dot{x} \approx \hat{f}_{M}(x) = A r(x)$, where $A \in \mathbb{R}^{n\times M}$ is a solution of
\begin{equation}
    A G_r = D.
\end{equation}

The use of $\hat{f}_{M}$ as an estimate of $f$ is justified by the following result, which follows from the fact that $P_{r} A_{f} P_{d}$ converges to $ A_{f} $ in norm as $M\to\infty$.
\begin{theorem}\label{thm:uniform-convergence}
    If $\mu_r < \mu_d$, $H_d = F^2_{\mu_d}(\mathbb{R}^n)$, $H_r = F^2_{\mu_r}(\mathbb{R}^n)$, $f:\mathbb{R}^n \to \mathbb{R}^n$ is a component-wise polynomial function, and the spans of the collections $ \{d_i\}_{i=1}^\infty $ and $\{r_i\}_{i=1}^\infty$ are dense in $H_{d}$ and $H_{r}$, respectively, then $\lim_{M\to\infty}\left(\sup_{x\in X}\left\Vert \hat{f}_{M}(x) - f(x) \right\Vert_{2}\right) = 0$.
\end{theorem}
\begin{pf}
   Since the space $F^2_{\mu_d}(\mathbb{R}^n)$ contains $(h_{\mathrm{id}})_j$ for $j=1,\cdots,n$, the functions $\hat{f}_{M,j} \coloneqq P_{r} A_{f} P_{d} (h_{\mathrm{id}})_j $ and $f_{j} \coloneqq A_{f} (h_{\mathrm{id}})_j $ that denote the $j-$th row of $\hat{f}_{M}$ and $f$, respectively, exist as members of $H_{r}$. Since  $x\mapsto K_r(x,x) = \exp\left(\frac{x^\top x}{\mu_r}\right)$ is continuous and $X$ is compact, there exists a real number $\overline{K}$ such that $\sup_{x\in X}K_r(x,x) = \overline{K}$. Since $\mu_r < \mu_d$, Theorem \ref{thm:compactness} implies that $A_f:F^2_{\mu_d}(\mathbb{R}^n)\to F^2_{\mu_r}(\mathbb{R}^n)$ is compact. Proposition \ref{prop:convergence} can then be used to conclude that for all $\epsilon > 0$ and $j=1,\ldots,n$, there exists $M(j) \in \mathbb{N}$ such that for all $i\geq M(j)$, $\left\Vert \hat{f}_{i,j} - f_{j}\right\Vert_{H_{r}}^{2} \leq \frac{\epsilon^2}{n\overline{K}^2}$. Using the reproducing property, for $i\geq \overline{M} \coloneqq \max_j M(j)$, 
    \begin{multline*}
        \left\Vert \hat{f}_{i}(x) - f(x) \right\Vert_{2}^{2} = \sum_{j=1}^{n} \left\langle \left(\hat{f}_{i,j} - f_{j}\right),K_r(\cdot,x)\right\rangle_{H_{r}}^{2}
        \\\leq \sum_{j=1}^{n} \left\Vert \hat{f}_{i,j} - f_{j} \right\Vert_{H_{r}}^{2}\left\Vert K_r(\cdot,x) \right\Vert_{H_{r}}^{2}\\
        \leq \sum_{j=1}^{n} \frac{\epsilon^2}{n\overline{K}^2}\left\langle K_r(\cdot,x),K_r(\cdot,x) \right\rangle_{H_{r}}^{2} = \frac{\epsilon^2}{\overline{K}^2}K_r(x,x)^{2}.
    \end{multline*}
    As a result, for all $\epsilon \geq 0$  there exists $\overline{M}$ such that for all $i\geq \overline{M}$,
    \begin{equation*}
        \sup_{x\in X}\left\Vert \hat{f}_{i}(x) - f(x) \right\Vert_{2} \leq \sqrt{\frac{\epsilon^2}{\overline{K}^2}\sup_{x\in X}K_r(x,x)^{2}} = \epsilon,
    \end{equation*}
    which completes the proof.\ifacqed
\end{pf}
This convergence result, typically unattainable for Koopman-based dynamic mode decomposition, is a key feature of this algorithm. 
\section{Occupation kernel regression} \label{sec:OKR}
Another approach to modeling is to directly use the occupation kernels as basis functions for regression.

If the components of the vector field $f$ reside in the RKHS, i.e., $f = \begin{bmatrix}
    f_1 & \ldots & f_n
\end{bmatrix}^{\top}$, with $f_i \in H$ for $i=1,\ldots,n$, then using the defining characteristic of the occupation kernel, the inner product $\langle f_i, \Gamma_{\gamma} \rangle_H$ may be expressed as
\[
    \langle f_i, \Gamma_{\gamma} \rangle_H = \int_0^T f_i(\gamma(t)) \mathrm{d}t = (\gamma(T))_i - (\gamma(0))_i
\]
for each $i=1,\ldots,n$. As such, given a set of solutions $\{ \gamma_{j}:[0,T_j] \to \mathbb{R}^n\}_{j=1}^M$, of $\dot{x} = f(x)$, components of the function $f$ can be estimated by minimizing the error between the inner products $\langle f_i, \Gamma_{\gamma} \rangle_H$ and the displacement $(\gamma(T))_i - (\gamma(0))_i$ of the $i-$th component of the trajectory. 

A regularized regression problem to determine an approximation $\hat f_i$ of the $i$-th row of $f$ can thus be formulated as
\begin{equation} \label{eq:RegularizedRegression}
    \min_{\hat f_i \in H} \sum_{j=1}^M \left( \langle \hat f_i, \Gamma_{\gamma_j} \rangle_H - \left( \gamma_{j}(T_j)\right)_i - 
    \left(\gamma_{j}(0)\right)_i\right)^2 + \lambda \| \hat f_i \|^2_H,
\end{equation}
where $\lambda > 0$ is a regularization parameter. Using the Representer theorem for occupation kernels \cite[Proposition 1]{SCC.Li.Rosenfeld2020}, the minimizer of \eqref{eq:RegularizedRegression} may be expressed as a linear combination of occupation kernels \[
    \hat f_i = A_{1,i} \Gamma_{\gamma_{1}}+ \cdots + A_{M,i} \Gamma_{\gamma_{M}} = A_i r(x),
\]
where $A_i \coloneqq \begin{bmatrix} A_{1,i}&\ldots&A_{M,i} \end{bmatrix}$. 
Using this representation of the minimizer, the inner products and the norm in the optimization problem can be computed as
\[
    \langle \hat{f}_i, \Gamma_{\gamma_j} \rangle_H =\left\langle \sum_{k=1}^M A_{k,i} \Gamma_{\gamma_{k}}, \Gamma_{\gamma_j} \right\rangle_H
    =
    (G_{r}^{j})^\top A_i^\top
\]
and
\begin{equation*}
    \| \hat f_i \|^2_H = A_i G_r A_i^\top,
\end{equation*}
where $G_{r}^j$ denotes the $j-$th column of $G_r$.

Hence, the resolution \eqref{eq:RegularizedRegression} reduces to the finite dimensional convex optimization problem
\begin{equation} \label{eq:RegularizedRegressionFinite}
    \min_{A_i \in \mathbb{R}^{1\times M}} \sum_{j=1}^M \left( (G_{r}^j)^\top A_i^\top - \left( \gamma_{j}(T_j) - 
    \gamma_{j}(0)\right)_i\right)^2 + \lambda A_i G_r A_i^\top.
\end{equation}
Solutions of the finite dimensional optimization problem coincide with solutions of the linear system $A_i(G_r + \lambda I_{M})G_r  = D_i G_r$, where $D_i$ is the $i-$th row of $D$ and $I_{M}$ denotes the $M\times M$ identity matrix.

Using the fact that the solution of $A_i(G_r + \lambda I_{M})  = D_i $ is one of the solutions of $A_i(G_r + \lambda I_{M})G_r  = D_i G_r$ (the only one if $G_r$ is nonsingular), the estimated model can be expressed as $\dot{x} \approx A r(x)$, where $A\in\mathbb{R}^{n\times M}$ is the solution of
\begin{equation}
     \label{eq:OKR_linear_system}
    A (G_r + \lambda I_{M}) = D.
\end{equation}

Interestingly, the model obtained via OKR coincides with that obtained using DMD if the regularization parameter is set to zero. As such, if occupation kernels are used for regression, the resulting model encodes additional structure, i.e., the singular functions and the singular values of the underlying Liouville operator, as provided by SLDMD. Furthermore, theorem \ref{thm:uniform-convergence}, also applicable to the regression model without regularization, provides an alternative to the cumulative prediction error convergence guarantees (see, for example \cite{SCC.Zhdanov.Kalnishkan2013} and Chapter 6 of \cite{SCC.Steinwart.Christmann2008}) that are typically available for regression problems.

\section{Numerical Experiments}\label{sec:sim}
The purpose of the numerical experiment is to demonstrate the effectiveness of the SLDMD algorithm, the OKR algorithm, and to compare the two methods. Additionally, the numerical experiment demonstrates the effects of the regularization constant $\lambda$ on the OKR method. 

This experiment utilizes the nonlinear model of the Duffing oscillator given by 
\begin{equation}
\label{Van_der_Pole}
\begin{array}{l}
\dot{x}_{1}=x_{2}, \\
\dot{x}_{2}=(x_{1}-x_{1}^3).
\end{array}
\end{equation}

To approximate the system dynamics for any given value of $\lambda>0$, 169 trajectories of the system are recorded starting from a grid of initial conditions. In the first trial, each trajectory is corrupted by Gaussian measurement noise with standard deviation 0.001. The initial velocities are obtained by numerically differentiating the measured noisy trajectories. In the second trial, the trajectories are not corrupted by noise, and the initial velocities are similarly obtained  by numerically differentiating the measured trajectories. The recorded trajectories are then utilized to approximate $f$. The SLDMD algorithm from Section \ref{sec:SLDMD} is implemented using the Moore-Penrose pseudoinverse. The OKR algorithm from Section \ref{sec:OKR} is implemented using the usual matrix inverse. Both methods use the kernel $K(x,y) = \exp\left(\frac{x^\top y}{\mu}\right)$ with $\mu=5$.


\begin{figure}
   \centering
    \input{figures/Duffing_COKR_f_tilde_changing_lambda_noise_in_signal}
    \caption{The blue marks are the mean$\left(  \left\vert \tilde{f}(x) \right\vert \right)$ over $x\in[-3,3]$ for different values of $\lambda$ using OKR trained with noisy trajectories. the dashed red line is the mean$\left(  \left\vert \tilde{f}(x) \right\vert \right)$ over $x\in[-3,3]$ calculated using the SLDMD method trained with noisy trajectories.} \label{fig:Duffing_COKR_f_tilde_changing_lambda_noise_in_signal}
\end{figure}
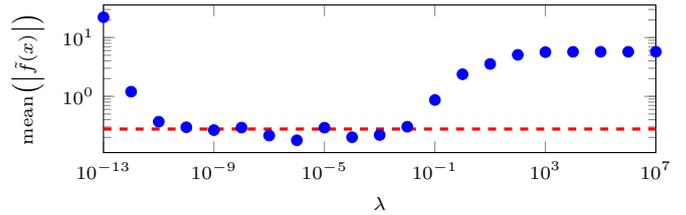

\begin{figure}
   \centering
    \input{figures/Duffing_COKR_f_tilde_changing_lambda_no_noise_in_signal}
    \caption{The blue marks are the mean$\left(  \left\vert \tilde{f}(x) \right\vert \right)$ over $x\in[-3,3]$ for different values of $\lambda$ using OKR trained with noise free trajectories. the dashed red line is the mean$\left(  \left\vert \tilde{f}(x) \right\vert \right)$ over $x\in[-3,3]$ calculated using the SLDMD method trained with noise free trajectories.} \label{fig:Duffing_COKR_f_tilde_changing_lambda_no_noise_in_signal}
\end{figure}
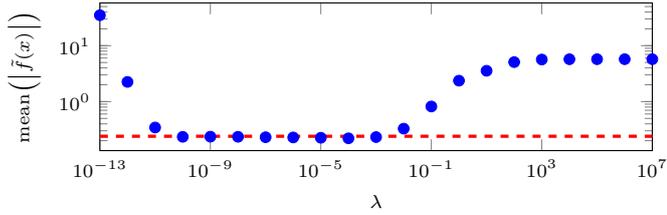

Figures \ref{fig:Duffing_COKR_f_tilde_changing_lambda_noise_in_signal} and \ref{fig:Duffing_COKR_f_tilde_changing_lambda_no_noise_in_signal}  show the mean of the difference between the actual values $f_2(x_1)$ and the estimated values $\hat{f_2}(x_1)$ over $x_2\in[-3,3]$ using different values of $\lambda$. Figure \ref{fig:Duffing_COKR_f_tilde_changing_lambda_noise_in_signal} shows the results for the system identification methods using noisy trajectories, where OKR clearly outperforms SLDMD for specific values of $\lambda$. Whereas, in Figure \ref{fig:Duffing_COKR_f_tilde_changing_lambda_no_noise_in_signal}, which shows the results of the two system identification methods using noise free trajectories, the OKR technique performs almost identically to the SLDMD technique for $\lambda$ values between $10^{-10}$ and $10^{-3}$. 

If the Gram matrix is not full rank, which is the case for both trials of the numerical experiments, then inversion of $G_r + \lambda I_{M}$ is numerically unstable for small values of $\lambda$. As such, consistent with Figures \ref{fig:Duffing_COKR_f_tilde_changing_lambda_noise_in_signal} and \ref{fig:Duffing_COKR_f_tilde_changing_lambda_no_noise_in_signal}, it is expected that the approximation $\hat{f}$ computed using OKR with a small value of $\lambda$ would be poor. On the other hand, the regularized inverse $(G_r + \lambda I_{M})^{-1}$ converges to the zero matrix as $\lambda$ increases, which explains the error plateau seen in Figures \ref{fig:Duffing_COKR_f_tilde_changing_lambda_noise_in_signal} and \ref{fig:Duffing_COKR_f_tilde_changing_lambda_no_noise_in_signal} for $\lambda>10^{2}$.

In theory, in the case where the trajectories are noise free and the inner products in the Gram matrix $G_r$ are evaluated exactly, the SLDMD algorithm should outperform OKR for any value of $\lambda$, since the regularization introduces a bias. The authors speculate that the slight improvement, seen in Figures \ref{fig:Duffing_COKR_f_tilde_changing_lambda_noise_in_signal} and \ref{fig:Duffing_COKR_f_tilde_changing_lambda_no_noise_in_signal} for $10^{-10} < \lambda < 10^{-3}$ is due to the integration errors introduced when computing the entries of the Gram matrix using Simpson's rule. Since regularization can prevent over-fitting when the underlying data are noisy, a more accurate approximation may be obtained for some values of $\lambda$. In the case where the trajectories are corrupted with measurement noise, there are errors in both $G_r$ and $D$ of (\ref{eq:OKR_linear_system}), which make the effects of $\lambda$ even more significant, consistent with Figure \ref{fig:Duffing_COKR_f_tilde_changing_lambda_noise_in_signal}.

\section{Conclusion}
This paper introduces a novel approach towards the construction of a finite rank representation of the Liouville operator. New results on the construction of singular values and functions of the finite rank operator using singular values and vectors of a matrix representation are also obtained. The singular values and functions give rise to a new dynamic mode decomposition model that is shown to be equivalent to regression using occupation kernels.

Numerical experiments that study the effect of the regularization parameter indicate that regularization may yield better results when the data are corrupted by integration errors and measurement noise. In order to support this conjecture a detailed Monte Carlo simulation is needed, which is a part of future work. 

\bibliography{scc,sccmaster,scctemp}

\end{document}

%% file: figures/Duffing_COKR_f_tilde_changing_lambda_noise_in_signal.tex
\begin{tikzpicture}
    \begin{loglogaxis}[
        xlabel={ $\lambda$ },
        ylabel={  mean$  \left(  \left\vert \tilde{f}(x) \right\vert \right) $},
        legend pos = outer north east,
        legend style={nodes={scale=0.5, transform shape}},
        enlarge y limits=0.1,
        enlarge x limits=0,
        height = 0.4\columnwidth,
        width = \columnwidth,
        label style={font=\scriptsize},
        tick label style={font=\scriptsize}
    ]
        \addplot [only marks, blue] table [x index=0, y index=1]{results/Duffing_COKR_f_tilde_changing_lambda_noise_in_signal.dat};

        \addplot [dashed, red, very thick] table [x index=0, y index=2]{results/Duffing_COKR_f_tilde_changing_lambda_noise_in_signal.dat};
        
    \end{loglogaxis}
\end{tikzpicture}

%% file: figures/Duffing_COKR_f_tilde_changing_lambda_no_noise_in_signal.tex
\begin{tikzpicture}
    \begin{loglogaxis}[
        xlabel={ $\lambda$ },
        ylabel={  mean$  \left(  \left\vert \tilde{f}(x) \right\vert \right) $},
        legend pos = outer north east,
        legend style={nodes={scale=0.5, transform shape}},
        enlarge y limits=0.1,
        enlarge x limits=0,
        height = 0.4\columnwidth,
        width = \columnwidth,
        label style={font=\scriptsize},
        tick label style={font=\scriptsize}
    ]
        \addplot [only marks, blue] table [x index=0, y index=1]{results/Duffing_COKR_f_tilde_changing_lambda_no_noise_in_signal.dat};

        \addplot [dashed, red, very thick] table [x index=0, y index=2]{results/Duffing_COKR_f_tilde_changing_lambda_no_noise_in_signal.dat};
        
    \end{loglogaxis}
\end{tikzpicture}